\documentclass[aps,prl,floatfix,showpacs,twocolumn,longbibliography,superscriptaddress]{revtex4-1}
\usepackage{ulem}
\usepackage{dcolumn}
\usepackage{bm}
\usepackage{color}
\usepackage{amssymb}
\usepackage{amsmath}
\usepackage{graphicx}
\usepackage{amsfonts}
\usepackage{slashed}
\usepackage{pstricks}
\usepackage{float}
\usepackage{multirow}
\usepackage{array}
\allowdisplaybreaks
\usepackage{dcolumn}
\usepackage{epsf}
\usepackage{float}
\usepackage[nice]{nicefrac}

\begin{document}
\title{Variational Monte Carlo calculations of $\mathbf{A\leq 4}$ nuclei with an artificial neural-network correlator ansatz}

\author{Corey Adams}
\affiliation{Physics Division, Argonne National Laboratory, Argonne, IL 60439}
\affiliation{Leadership Computing Facility, Argonne National Laboratory, Argonne, IL 60439}

\author{Giuseppe Carleo}
\affiliation{Institute of Physics, École Polytechnique Fédérale de Lausanne (EPFL), CH-1015 Lausanne, Switzerland}

\author{Alessandro Lovato}
\affiliation{Physics Division, Argonne National Laboratory, Argonne, IL 60439}
\affiliation{INFN-TIFPA Trento Institute of Fundamental Physics and Applications, 38123 Trento, Italy}

\author{Noemi Rocco}
\affiliation{INFN-TIFPA Trento Institute of Fundamental Physics and Applications, 38123 Trento, Italy}
\affiliation{Theoretical Physics Department, Fermi National Accelerator Laboratory, P.O. Box 500, Batavia, Illinois 60510, USA}

\date{\today}
\begin{abstract}
The complexity of many-body quantum wave functions is a central aspect of several fields of physics and chemistry where non-perturbative interactions are prominent. Artificial neural networks (ANNs) have proven to be a flexible tool to approximate quantum many-body states in condensed matter and chemistry problems. In this work we introduce a neural-network quantum state ansatz to model the ground-state wave function of light nuclei, and approximately solve the nuclear many-body Schr\"odinger equation.   
Using efficient stochastic sampling and optimization schemes, our approach extends pioneering applications of ANNs in the field, which present exponentially-scaling algorithmic complexity. We compute the binding energies and point-nucleon densities of $A\leq 4$ nuclei as emerging from a leading-order pionless effective field theory Hamiltonian. We successfully benchmark the ANN wave function against more conventional parametrizations based on two- and three-body Jastrow functions, and virtually-exact Green's function Monte Carlo results. 
\end{abstract} 
\maketitle

\textit{Introduction --} 
The last two decades have witnessed remarkable progress in our understanding of how the structure and dynamics of atomic nuclei emerge from the individual interaction among protons and neutrons. This progress has been primarily driven by the widespread use of nuclear-effective field theories to systematically construct realistic Hamiltonians~\cite{Epelbaum:2008ga,Machleidt:2011zz,Hammer:2019poc}, and the concurrent development of nuclear many-body techniques that solve the time-independent Schr\"odinger equation with controlled approximations~\cite{Barrett:2013nh,Hagen:2013nca,Hergert:2015awm,Carbone:2013eqa,Epelbaum:2011md}. The variational Monte Carlo (VMC) and the Green's Function Monte Carlo (GFMC) methods are ideally suited to tackle this problem and have been extensively applied to study properties of light nuclei~\cite{Carlson:2014vla}. { Monte Carlo techniques also face important challenges. For example,} the calculation of the spin-isospin dependent Jastrow correlations used in the VMC and GFMC scales exponentially with the number of nucleons, limiting the applicability of these methods to relatively small nuclear systems. Also, the auxiliary-field diffusion Monte Carlo~\cite{Schmidt:1999lik} (AFDMC) samples the spin and isospin degrees of freedom to treat larger nuclei and infinite nucleonic matter~\cite{Piarulli:2019pfq,Lonardoni:2019ypg}, but it can only  take as inputs somewhat simplified interactions~\cite{Gandolfi:2020pbj}. In addition, the use of wave functions that scale polynomially with the number of nucleons exacerbates the AFDMC fermion sign problem for $A > 16$ nuclei. { Therefore, extending VMC and GFMC calculations to medium-mass nuclei requires devising accurate wave functions that exhibit a polynomial scaling with A.}

An alternative class of approaches being actively explored is based on machine learning (ML) techniques. These techniques typically rely on the ability of artificial neural networks (ANNs) to compactly represent complex high-dimensional functions, as already leveraged in several domains of physics \cite{carleo_machine_2019}. For many-body quantum applications, neural-network-based variational representations have been introduced in Ref.~\cite{Carleo:2017}, and have found applications as a tool to study ground state and dynamics of several interacting lattice quantum systems ~\cite{nomura_restricted_2017,Saito:2018b,Choo:2018,Nomura:2020,yoshioka:2019,nagy_variational_2019,vicentini:2019,hartmann_neural-network_2019,ferrari_neural_2019}.
In a series of recent works~\cite{Pfau:2019,Hermann:2019,Choo:2019} deep neural networks have been further developed to tackle ab-initio chemistry problems within variational Monte Carlo, often resulting in accuracy improvements over existing variational approaches traditionally used to describe correlated molecules. 
While applications of ML approaches to the many-body problem in condensed matter, quantum chemistry, and quantum information have been proliferating in the past few years, the adoption in low-energy nuclear theory is still in its infancy~\cite{Negoita:2018kgi,Jiang:2019zkg}. Pioneering work in the field ~\cite{Keeble:2019bkv} has provided a proof-of-principle application of ANN to solve the Schr\"odinger equation of the deuteron. Extending the non-stochastic approach of Ref.~\cite{Keeble:2019bkv} to larger nuclei however presents an intrinsically exponentially scaling challenge. 

In this work, we expand the domain of applicability of ANN-based representations of the wave function and compute ground-state properties of $A\leq 4$ nuclei as they emerge from a leading-order pionless effective field theory (EFT) Hamiltonian, containing consistent two- and three-body potentials. { Specifically, we develop a novel VMC algorithm based on an ANN representation of the spin-isospin dependent correlator that captures the vast majority of nuclear correlations and scales favorably with the number of nucleons.} We benchmark our results against a more conventional parametrization of the variational wave function in terms of two- and three-body Jastrow functions, and virtually-exact GFMC calculations.

\paragraph{Hamiltonian.-} We employ nuclear Hamiltonians derived within pionless EFT, which is based on the tenet that the typical momentum of nucleons in nuclei is much smaller than the pion mass $m_\pi$~\cite{Bedaque:2002mn,Hammer:2019poc}. Under this assumption, largely justified for studying the structure and long-range properties of $A\leq 4$ nuclei, pion exchanges are unresolved contact interactions and nucleons are the only relevant degrees of freedom. The singularities of the contact terms are controlled introducing a Gaussian regulator that suppresses transferred momenta above the ultraviolet cutoff $\Lambda$. This regulator choice directly leads to a Gaussian radial dependence of the potential, which is local in coordinate~\cite{Barnea:2013uqa,Contessi:2017rww}. The leading-order (LO) Hamiltonian reads
\begin{align}
H_{LO} &=-\sum_i \frac{{\vec{\nabla}_i^2}}{2m_N} 
+\sum_{i<j} {\left(C_1  + C_2\, \vec{\sigma_i}\cdot\vec{\sigma_j}\right) 
e^{-r_{ij}^2\Lambda^2 / 4 }}
\nonumber\\
&+D_0 \sum_{i<j<k} \sum_{\text{cyc}} 
{e^{-\left(r_{ik}^2+r_{ij}^2\right)\Lambda^2/4}}\,,
\label{eq:ham}
\end{align}
where $m_N$ is the mass of the nucleon, $\vec{\sigma_i}$ is the Pauli matrix acting on nucleon $i$, and $\sum_{\text{cyc}}$ stands for the cyclic permutation of $i$, $j$, and $k$. 

Following Ref.~\cite{Kirscher:2015yda}, the low-energy constants $C_1$ and $C_2$ are fit to the deuteron binding energy and to the neutron-neutron scattering length. In Eq. \eqref{eq:ham} we picked the operator basis $1$ and $\vec{\sigma_i}\cdot\vec{\sigma_j}$, but this choice can be replaced by any other form equivalent under Fierz transformations in SU(2). Solving $A\geq 3$ nuclei with purely attractive two-nucleon potentials leads to the ``Thomas collapse''~\cite{Yang:2019hkn}, which can be avoided promoting a contact three-nucleon force to LO~\cite{Bedaque:1998kg}. 
The values of the LECs’ adopted in this work can be found in Ref.~\cite{Kirscher:2015yda}; since $C_1(\Lambda)$ is much larger than $C_2(\Lambda)$, the LO Hamiltonian has an approximate SU(4) symmetry. 

\paragraph{Variational wave function.-} 

A fundamental ingredient of the VMC method is the choice of a suitable variational wave function $\Psi_V$, whose parameters are found exploiting the variational principle 
\begin{align}
\frac{\langle \Psi_V | H | \Psi_V \rangle}{\langle \Psi_V | \Psi_V \rangle} = E_V \geq E_0 
\label{eq:H_exp}
\end{align}
where $E_0$ is exact the ground-state energy: $H|\Psi_0\rangle = E_0 |\Psi_0\rangle$. The Metropolis Monte Carlo algorithm is used to evaluate {the variational energy} $E_V$ by sampling the spatial and spin-isospin coordinates. { We introduce the following ANN representation of the variational wave function
\begin{align}
|\Psi_V^{\textrm{ANN}} \rangle =  e^{\,\mathcal{U}(\mathbf{r}_1,\dots,\mathbf{r}_A)} \tanh[\mathcal{V}(\mathbf{s}_1,\mathbf{r}_1,\dots,\mathbf{r}_A, \mathbf{s}_A)]  |\Phi\rangle\,  
\label{eq:psi_ANN}
\end{align}
where $\{\mathbf{r}_1,\dots,\mathbf{r}_A\}$ and $\{\mathbf{s}_1,\dots,\mathbf{s}_A\}$ denote the set of single-particle spatial three-dimensional coordinates and the z-projection of the spin-isospin degrees of freedom $\mathbf{s}_i = \{s^z_i, t_i^z\}$, respectively. For the s-shell nuclei considered in this work, we take $|\Phi_{^2\rm H}\rangle = \mathcal{A} |\uparrow_p \uparrow_n \rangle$, $|\Phi_{^3\rm He}\rangle = \mathcal{A} |\uparrow_p \downarrow_p  \uparrow_n \rangle$, and $|\Phi_{^4\rm He} \rangle = \mathcal{A} |\uparrow_p \downarrow_p  \uparrow_n \downarrow_n \rangle$, with $\mathcal{A}$ being the anti-symmetrization operator~\cite{Lomnitz-Adler:1981dmh}.

The real-valued correlating factors $\mathcal{U}(\mathbf{r}_1,\dots,\mathbf{r}_A)$ and 
and $\mathcal{V}(\mathbf{s}_1,\mathbf{r}_1,\dots,\mathbf{r}_A, \mathbf{s}_A)$ are parametrized in terms of permutation-invariant ANNs, so that the total wave function is anti-symmetric. To achieve this goal, we make use of the Deep Sets architecture~\cite{Zaheer:2017, Wagstaff:2019}, and map each of the single-particle inputs separately to a latent-space representation. We then apply a sum operation to destroy the ordering of the information and ensure permutation invariance
\begin{equation}
\mathcal{F}(\mathbf{x}_1,\dots,\mathbf{x}_A) = \rho_\mathcal{F}\left(\sum_{\mathbf{x}_i} \phi_\mathcal{F}(\mathbf{x}_i)\right)\, \quad \mathcal{F} = \mathcal{U}, \mathcal{V}\, .
\end{equation}
Both $\phi_\mathcal{U}$ and $\rho_\mathcal{U}$ are represented by ANNs comprised of four fully connected layers with 32 nodes each, while $\phi_\mathcal{V}$ and $\rho_\mathcal{V}$ are made of two fully connected layers, again with 32 nodes, for total of $13058$ trainable parameters. The calculation of the kinetic energy requires using differentiable activation functions. We find that $\tanh$ and softplus~\cite{Dugas:2000} yield fully consistent results. The single-particle inputs are $\mathbf{x}_i \equiv \{\bar{\mathbf{r}}_i\}$ and $\mathbf{x}_i \equiv \{\bar{\mathbf{r}}_i, \mathbf{s}_i\}$ for $\mathcal{U}$ and $\mathcal{V}$, respectively, where we defined intrinsic spatial coordinates as $\bar{\mathbf{r}}_i = \mathbf{r}_i - \mathbf{R}_{\rm CM}$, with $\mathbf{R}_{\rm CM}$ being the center of mass coordinate. This procedure automatically removes spurious center of mass contributions from all observables~\cite{Massella:2018xdj}.} Since the parameters of the network are randomly initialized, in the initial phases of the training, during the Metropolis walk, the nucleons can drift away from $\mathbf{R}_{\rm CM}$. To control this behavior, a Gaussian function is added to confine the nucleons within a finite volume $\mathcal{U}(\mathbf{r}_1,\dots,\mathbf{r}_A) \to \mathcal{U}(\mathbf{r}_1,\dots,\mathbf{r}_A) -\alpha \sum_i \bar{\mathbf{r}}_i^2$ where we take $\alpha = 0.05$.


The choice of correcting a mean-field state $|\Phi\rangle $ with a flexible ANN correlator factor is similar in spirit to neural-network correlators introduced recently in condensed-matter \cite{nomura_restricted_2017,ferrari_neural_2019} and chemistry applications \cite{Hermann:2019}{, but it is more general as it encompasses spin-isospin dependent correlations}. An appealing feature of the ANN ansatz is that it is more general than the more conventional product of two- and three-body spin-independent Jastrow functions
\begin{align}
|\Psi_V^J \rangle = \prod_{i<j<k} \Big( 1-\sum_{\text{cyc}} u(r_{ij}) u(r_{jk})\Big) \prod_{i<j} f(r_{ij}) | \Phi\rangle\,,
\label{eq:psi_T}
\end{align}
which is commonly used for nuclear Hamiltonians that do not contain tensor and spin-orbit terms~\cite{Contessi:2017rww, Schiavilla:2021dun}. 

\begin{figure}[b]
    \centering
    \includegraphics[width=\columnwidth]{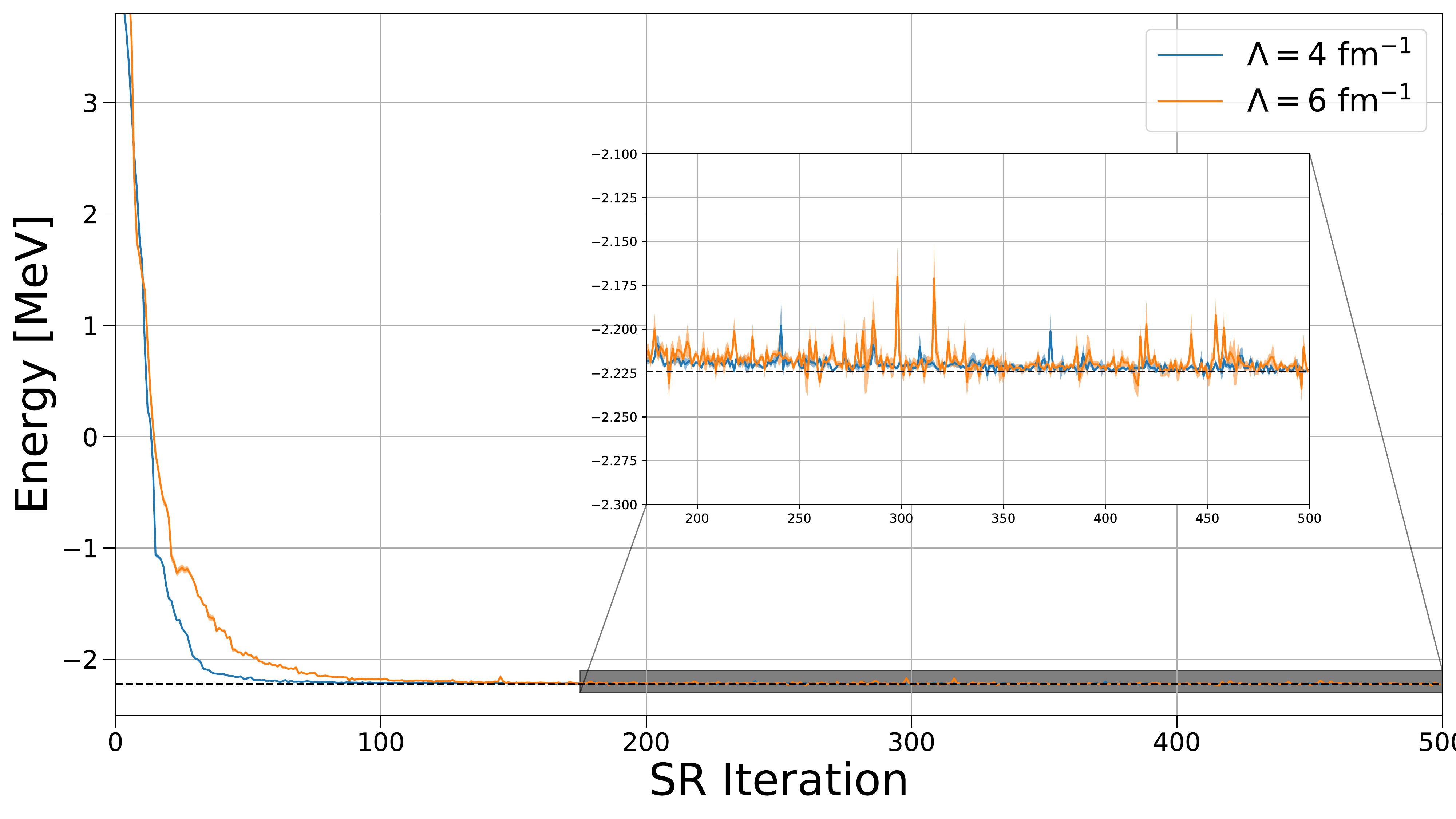}
    \caption{Convergence pattern of the $^2$H variational energy for  $\Lambda = 4$ fm$^{-1}$ and $\Lambda = 6$ fm$^{-1}$ as a function of the number of optimization steps of the SR \textit{AdaptiveEta} algorithm. The dashed line denotes the asymptotic value.}
    \label{fig:SR}
\end{figure}

Analogously to standard VMC calculations, as well as ML applications, the optimal set of weights and biases of the ANN is found minimizing a suitable cost function. Specifically, we exploit the variational principle of Eq.~\eqref{eq:H_exp} and minimize the expectation value of the energy. The gradient components $G_i=\partial_i E(p)$ of the energy with respect to the variational parameters $p_i$ read 
\begin{align}
    G_{i}=2\left( \frac{\langle\partial_{i}\Psi_V|H|\Psi_V\rangle}{\langle\Psi_V|\Psi_V\rangle}-E_V \frac{\langle\partial_{i}\Psi_V|\Psi_V\rangle}{\langle\Psi_V|\Psi_V\rangle}\right)
\end{align}
and can be efficiently estimated through Monte Carlo sampling. 
While stochastic gradient descent can be readily used to compute parameters updates, for VMC applications it has been found that using a preconditioner based on the Quantum Fisher Information 
\begin{align}
    S_{ij}=\frac{\langle\partial_{i}\Psi_V|\partial_{j}\Psi_V\rangle}{\langle\Psi_V|\Psi_V\rangle}-\frac{\langle\partial_{i}\Psi_V|\Psi_V\rangle\langle\Psi_V|\partial_{j}\Psi_V\rangle}{\langle\Psi_V|\Psi_V\rangle\langle\Psi_V|\Psi_V\rangle},
\end{align}
is significantly more efficient. During the optimization, then parameters at step $s$ are updated as $p^{s+1}=p^{s}-\eta (S+\Lambda)^{-1}G$, where $\eta$ is the learning rate and $\Lambda$ is a small positive diagonal matrix that is added to stabilize the method. This approach, known as the stochastic-reconfiguration (SR) algorithm~\cite{sorella_green_1998,Sorella:2005} is equivalent to performing imaginary-time evolution in the variational manifold and it is in turn related to the Natural Gradient descent method \cite{amari_natural_1998} in unsupervised learning.
{Our computational techniques are based on the general ML framework \texttt{Tensorflow} \cite{tensorflow2015-whitepaper}, and it is scalable across more than 100 GPUs. We also maintain an additional developmental repository written in \texttt{JAX} \cite{jax2018github} for fast prototyping of new features. More information about the architecture and performance of the software is available in the supplemental material.}

Figure~\ref{fig:SR} displays the convergence pattern as function of the optimization step of the $^{2}$H energy for the LO pionless EFT Hamiltonians with $\Lambda = 4$ fm$^{-1}$ and $\Lambda = 6$ fm$^{-1}$.
In the initial phase of the optimization, the softer cutoff exhibits a faster convergence than the stiffer one. However the asymptotic value of the energy is reached after about $300$ iterations for both values of the regulator. These results have been obtained using an adaptive learning rate in the range $10^{-7} \leq \eta \leq 10^{-2}$, which has proven to yield robust convergence patterns for all the nuclei and regulator choices that we have analyzed. The adaptive schedule of this \textit{AdaptiveEta} algorithm is selected performing heuristic tests on the parameter change, similar to the ones introduced in Ref.~\cite{Contessi:2017rww, Massella:2018xdj} for regularizing the linear optimization method~\cite{Toulouse:2007}.

\paragraph{Results and discussion. -} 
\begin{table}[t]
\renewcommand{\arraystretch}{1.1}
\begin{center}
\begin{tabular}{ c | c | c c c c c }
\hline
\hline
& $\Lambda$ & VMC-ANN  &  VMC-JS & GFMC & GFMC$_c$\\
\hline
\multirow{2}{*}{$^2$H } & $4$ fm$^{-1}$ & $-2.224(1)$  &   $-2.223(1)$  & $-2.224(1)$ & - \\
& $6$ fm$^{-1}$ & $-2.224(4)$  &   $-2.220(1)$  & $-2.225(1)$ & - \\
\hline
\multirow{2}{*}{$^3$H } & $4$ fm$^{-1}$ & $-8.26(1)$  &   $-7.80(1)$  & $-8.38(2)$ & $-7.82(1)$ \\
& $6$ fm$^{-1}$ & $-8.27(1)$  &   $-7.74(1)$  & $-8.38(2)$ &$-7.81(1)$ \\
\hline
\multirow{2}{*}{$^4$He } & $4$ fm$^{-1}$ & $-23.30(2)$  &   $-22.54(1)$  & $-23.62(3)$ & $-22.77(2)$ \\
& $6$ fm$^{-1}$ & $-24.47(3)$  &   $-23.44(2)$  & $-25.06(3)$ &$-24.10(2)$\\
\hline
\end{tabular}
\caption{Ground-state energies in MeV of the $^2$H, $^3$H, and $^4$He for the LO pionless-EFT Hamiltonian for $\Lambda = 4$ fm$^{-1}$ and $\Lambda = 6$ fm$^{-1}$. Numbers in parentheses indicate the statistical errors on the last digit.}
\vspace{-0.4cm}
\label{tab:2h}
\end{center}
\end{table}

We analyze the accuracy of the ANN wave function ansatz by computing the ground-state energies of $^2$H, $^3$H, and $^4$He. In Table~\ref{tab:2h} we benchmark the ANN representation of $\Psi_T$ (VMC-ANN) against conventional VMC calculations carried out using a spline parametrization for the Jastrow functions~\cite{Contessi:2017rww} (VMC-JS), and virtually-exact GFMC results. 

The three methods provide fully compatible energies for $^2$H nucleus, within statistical errors, showing the flexibility of the ANN to accurately represent the ground-state wave function of the deuteron, consistent with the findings of Ref.~\cite{Keeble:2019bkv}. Note that, since the LO pionless EFT Hamiltonian does not contain tensor or spin-orbit terms, the VMC-JS ansatz is exact. The perfect agreement with the experimental value is not surprising, as the potential has been fit to the deuteron binding energy using numerically-exact few-body methods~\cite{Barnea:2013uqa}.

The VMC-ANN noticeably improves upon the VMC-JS energies of $^3$H, by $\simeq 0.5$ MeV for both $\Lambda = 4$ fm$^{-1}$ and $\Lambda = 6$ fm$^{-1}$. On the other hand, the GFMC results are $\simeq 0.1$ MeV more bound than the VMC-ANN ones. This difference is due to spin-dependent correlations that are automatically generated by the GFMC imaginary-time propagation, but are not fully accounted for by the correlator ansatz of Eq.~\eqref{eq:psi_ANN}. To better quantify the spin-independent correlations entailed in the ANN, we have considered a simplified ``ANN$_c$'' ansatz $|\Psi_V^{\textrm{ANN}_c} \rangle =  e^{\,\mathcal{U}(\mathbf{r}_1,\dots,\mathbf{r}_A)} |\Phi\rangle$. In this case, the NN potential of Eq.~\eqref{eq:ham} is equivalent to the $SU(4)$-symmetric interaction $\tilde{v}_c(r_{ij}) = v_c(r_{ij}) - v_\sigma(r_{ij})$. For $\Lambda = 4$ fm$^{-1}$ and $\Lambda = 6$ fm$^{-1}$ ANN$_c$ yields $-7.85(2)$ and $-7.85(4)$ MeV, respectively. These numbers are in excellent agreement with the GFMC$_c$ calculations reported in Table~\ref{tab:2h}, which have also been carried out using $\tilde{v}_c(r_{ij})$.

A similar pattern emerges for $^4$He, with ANN wave functions outperforming the JS ones: the energy is improved by about $0.8$ MeV and $1.0$ MeV for $\Lambda = 4$ fm$^{-1}$ and $\Lambda = 6$ fm$^{-1}$, respectively. The small discrepancies with the GFMC are again due to missing spin-isospin dependent correlations in the ANN. In fact, the ANN$_c$ energies turn out to be $-22.76(2)$ MeV and $-24.05(5)$ for $\Lambda = 4$ fm$^{-1}$ and $\Lambda = 6$ fm$^{-1}$, which are fully compatible with the GFMC$_c$ results listed in Table~\ref{tab:2h}.

\begin{figure}[t]
    \centering
    \includegraphics[width=\columnwidth]{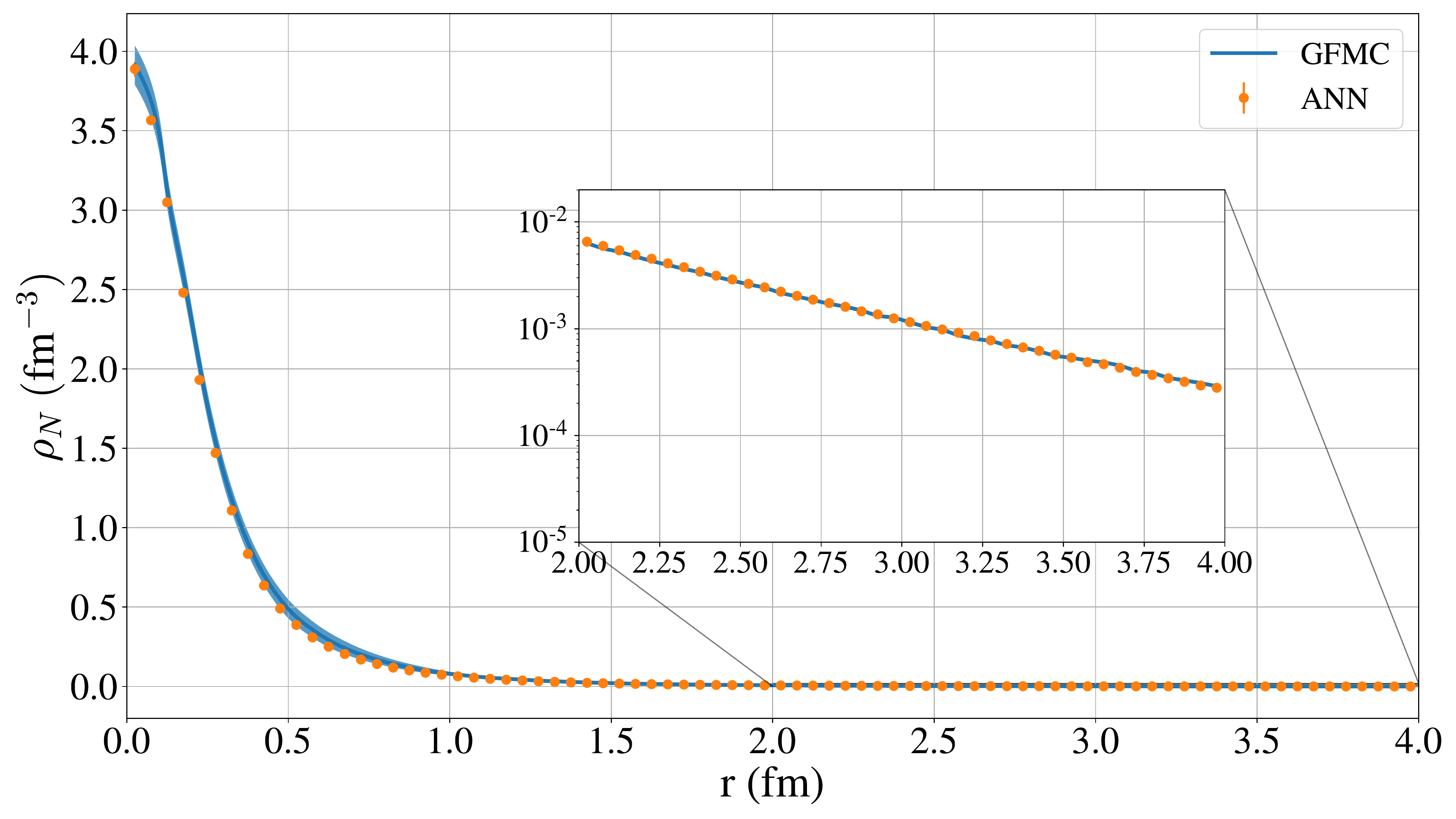}
    \includegraphics[width=\columnwidth]{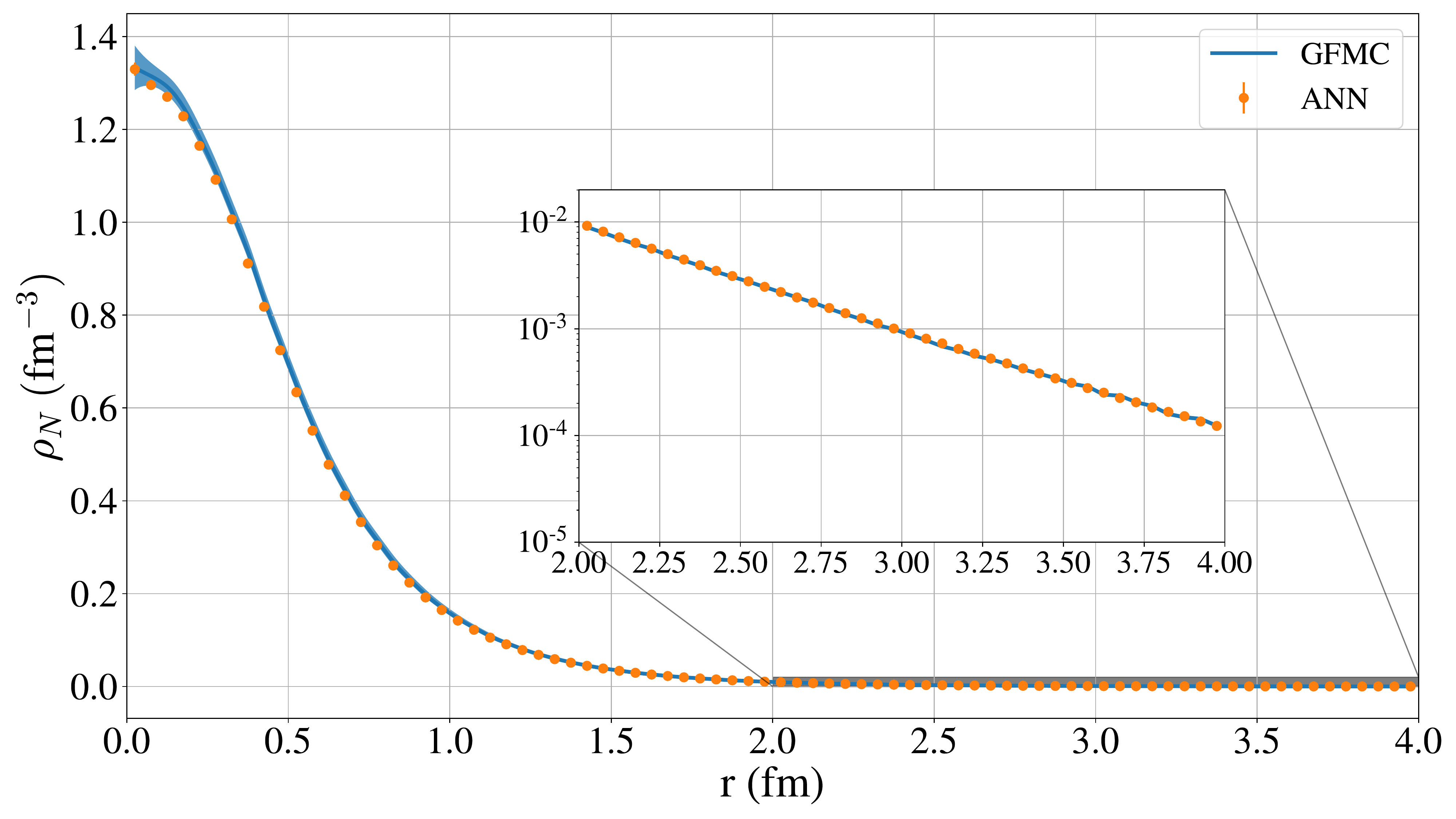}
    \includegraphics[width=\columnwidth]{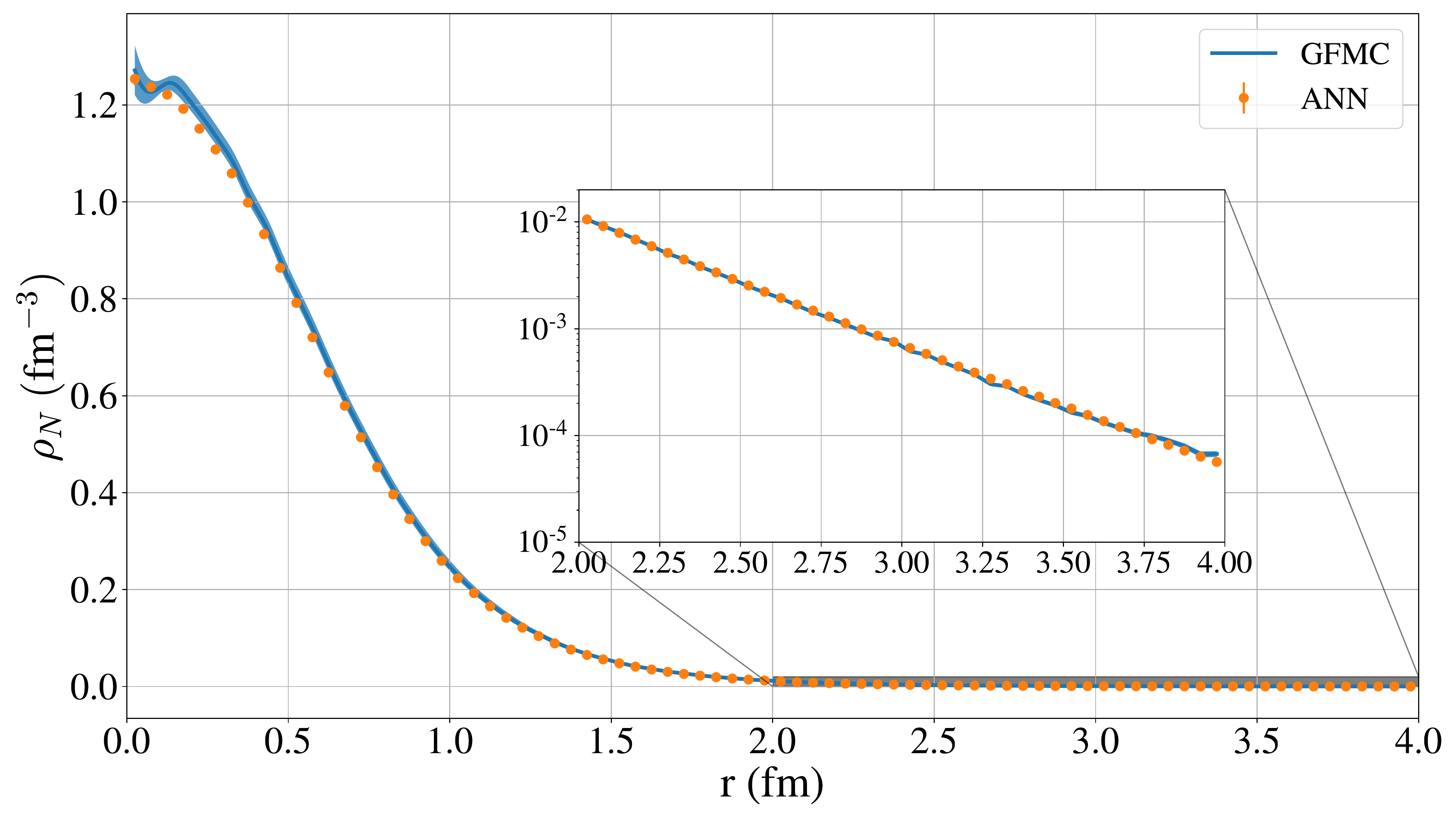}
    \caption{Point-nucleon densities of $^2$H (upper panel), $^3$H (middle panel), and $^4$He (lower panel) for the LO pionless-EFT Hamiltonian with $\Lambda =4$ fm$^{-1}$. The solid points and the shaded area represent the VMC-ANN and GFMC results, respectively.}
    \label{fig:rho_2h}
\end{figure}

To further elucidate the quality of the ANN wave function we consider the point-nucleon density
\begin{align}
	\rho_{N}(r) &=\frac{1}{4\pi r^2}\big\langle\Psi_V \big|\sum_i \mathcal \delta(r-|\mathbf{r}_i^{\rm int}|)\big|\Psi_V \big\rangle\,, 
	\label{eq:rho_N}
\end{align}
which is of interest in a variety of experimental settings~\cite{Lovato:2013cua,Weiss:2018zrd}. In the upper, medium, and lower panels of Fig.~\ref{fig:rho_2h} we display $\rho_{N}(r)$ of $^2$H, $^3$H, and $^4$He as obtained from VMC-ANN and GFMC calculations that use as input the LO pionless-EFT Hamiltonian with $\Lambda =4$ fm$^{-1}$. There is an excellent agreement between the two methods, which further corroborates the representative power of the ANN ansatz for the wave functions of $A\leq 4$ nuclei. The VMC-ANN and GFMC densities overlap both at short distances and in the slowly-decaying asymptotic exponential tails, highlighted in the insets of Fig.~\ref{fig:rho_2h}. It should be emphasised that the ANN learns how to compensate for the original Gaussian confining function and reproduce the correct exponential falls off of the nuclear wave function, which is notoriously delicate to obtain within nuclear methods that rely on harmonic-oscillator basis expansions~\cite{Vary:2018jsp, Sharaf:2019jgw}. 

\paragraph{Conclusions --} In this work we have carried out proof-of-principle calculations that demonstrate the capability of ANNs to represent the variational state of $A\leq 4$ nuclei encompassing the vast majority of nuclear correlations and scale favorably with the number of nucleons. Exploiting the Deep Sets architecture, we have devised permutation-invariant, spin-isospin dependent correlators whose computational cost scales polynomially with the number of nucleons. Using the stochastic-reconfiguration algorithm, we solve the Schr\"odinger equation of a LO pionless-EFT Hamiltonian that contains two- and three-nucleon potentials characterized by highly non-perturbative, spin dependent, short-range components. The spin-isospin dependent ANN variational wave function outperforms the routinely employed two- and three-body Jastrow parametrization of the correlation function. The small remaining differences with the exact GFMC result will likely be solved once spin-dependent backflow correlations are introduced in the Slater determinant, as in Ref.~\cite{Pfau:2019,Hermann:2019,Choo:2019}, paving the way for performing accurate quantum Monte Carlo studies of medium-mass nuclei.

The single-particle densities obtained with ANN wave functions are also in excellent agreement with GFMC results, both at short distances and in the slowly-decaying exponential tails, which are notoriously difficult to reproduce.

\paragraph{Acknowledgments -- } 
The present research is supported by the U.S. Department of Energy, Office of Science, Office of Nuclear Physics, under contracts DE-AC02-06CH11357, by the NUCLEI SciDAC program (A.L. and N.R.) and by Fermi Research Alliance, LLC under Contract No. DE-AC02-07CH11359 with the U.S. Department of Energy, Office of Science, Office of High Energy Physics (N.R.). This research used resources of the Argonne Leadership Computing Facility, which is a DOE Office of Science User Facility supported under Contract DE-AC02-06CH11357.  The calculations were performed using resources of the Laboratory Computing Resource Center of Argonne National Laboratory. We acknowledge discussions with Markus Holzmann, Dean Lee, James Stokes, and James Vary. 

\bibliography{biblio}

\end{document}